\documentclass[floatfix,showpacs,preprintnumbers,amsmath,amssymb]{revtex4}

\usepackage{graphicx}
\usepackage{dcolumn}
\usepackage{bm}
\def\be{\begin{equation}}
\def\ee{\end{equation}}
\def\bea{\begin{eqnarray}}
\def\eea{\end{eqnarray}}

\begin{document}
\title{The general relativistic geometry of the Navarro-Frenk-White model}
\author{Tonatiuh~Matos$^{1}$, and Dar\'{\i}o~N\'{u}\~{n}ez$^{2}$}
\affiliation{$^{1}$ Departamento de F{\'\i}sica, Centro de
Investigaci\'on y de Estudios Avanzados del IPN, A.P. 14-740,
07000 M\'exico D.F., M\'exico.\\$^{2}$Instituto de Ciencias
Nucleares - UNAM, Apdo. 70-543, Ciudad Universitaria, 04510
M\'exico, D.F., M\'exico}

\email{tmatos@fis.cinvestav.mx, nunez@nuclecu.unam.mx}

\begin{abstract}
We derive the space time geometry associated with the Navarro
Frenk White dark matter galactic halo model. We discuss several
properties of such a spacetime, with particular attention to the
corresponding Newtonian limit, stressing the qualitative and
quantitative nature of the differences between the relativistic
and the Newtonian description. We also discuss on the
characteristics of the possible stress energy tensors which could
produce such a geometry, via the Einstein's equations.
\end{abstract}

\date{\today}


\maketitle


The geometry generated by the galactic halo, is generally thought
to be "almost flat". This assumption is based, first, on the fact
that the galactic dark halo has very low density, at most a few
order of magnitude above the critical one. Second, the velocities
involved are small compared with the speed of light, and third,
the dust treatment gives a description of the dynamics which is in
good agrement with the observation. Thus, validates the Newtonian
physics as an adequate to treat the dark halo.

These very same arguments are used for studying the Solar system,
where also the geometry is taken as "almost flat". Nevertheless,
the general relativistic treatment of the Solar system has made
possible to give important corrections to the Newtonian one and,
furthermore, in the general relativistic treatment is precisely
this "almost not flatness", which explains the motion of the
planets!

We consider that to count with a general relativistic version of
the galactic dark matter halo allows one to make more accurate
analysis on the dynamics of the objects, including the study on
gravitational lenses, to mention an application.

In the present work, we describe how the observations can be
related with part of the geometry, then propose and expression for
the complete geometry associated with the Navarro-Frenk-White,
NFW, model. Next describe the properties of such a geometry and
explain why the Newtonian description works so well. Armed with
the geometry, we discuss on the type of matter-energy which
generates the geometry, that is, the nature of dark matter, a
point where the Newtonian analysis remains mute.

In a previous series of works, Matos et al.\cite{larg}, Guzm\'{a}n
et. al. \cite{quint}, we discussed the possibility of determining
the geometry of the space-time, and then constraining the type of
matter-energy which generate such geometry, based on observational
data. In particular, we addressed the problem of making those
determinations based on the observed profile of the tangential
velocities of objects orbiting galaxies.

In a nutshell: Given the fact that the dark halo in the galaxies
seems to be spherical and at rest, at least in the average, we
consider a general spherically symmetric static space-time, see
Eq.(\ref{eq:els}) below, and were able to determine, on purely
geometrical ground, an expression for the tangential velocity of
objects moving in circular stable geodesics in terms of the metric
coefficients, which turned out to be a very simple one,
Eq.(\ref{eq:velt}). We then took a sort of inverse point of view.
Instead of consider such an equation as an expression for the
velocity, we took it as an expression for the metric coefficient,
given the fact that what is being observed is the velocity
profile, thus being able to determine part of the geometry based
only on observational data, Eq.(\ref{eq:alph}).

In the present work we present such a program applied to the NFW
model, Navarro et al. \cite{NFW_mod}, which has proved to have a
remarkable predicted power and agrees very well with observations,
particularly with those outside the central galactic region.

In what follows we reproduce the main steps on the reasoning
leading to the conditions which the tangential velocity of
circular orbits impose on the metric coefficients for the
spherically symmetric static case.

We begin with the general line element for such geometry:
\begin{equation}
ds^2=-\alpha^2(r)\,c^2\,dt^2+\frac{dr^2}{\left(1-\frac{2\,G\,M(r)}{c^2\,r}\right)}+
r^2\,d\Omega^2, \label{eq:els}
\end{equation}
where $c$ is the speed of light and $G$ the gravitational
constant, and $d\Omega^2=d\theta^2+\sin^2\theta\,d\varphi^2$ is
the solid angle element. From the corresponding Lagrangian for a
test particle in this space,
$2\mathcal{L}=\left(\frac{ds}{d\tau}\right)^2$, where $\tau$
stands for the proper time, we obtain that the energy,
$E=\alpha^2(r)\,c^2\,\dot{t}$, the $\varphi $-momentum
$L_{\varphi}=r^2\,\sin^2\theta\dot{\varphi}$, and the total
angular momentum, $L^2={L_{\theta}}^2+
({\frac{{L_{\varphi}}}{{\sin \theta }}})^2$, with $L_{\theta
}=r^{2}\dot{\theta}$, where dot stands for derivative with respect
to the proper time, are conserved quantities along the motion.
Notice that we can write the total angular momentum in terms of
the solid angle as: $L^2=r^2\,{\dot{\Omega}}^2$.

With this information, the fact that the four-velocity,
$u^\mu=\frac{dx^\mu}{d\tau}$, is normalized, $u_\nu\,u^\nu=-1$,
translates into a radial motion equation:
\begin{equation}
\dot{r}^{2}+V(r)=0,  \label{eq:rades}
\end{equation}
with the potential $V(r)$ given by
\begin{equation}
V(r)=-\left(1-\frac{2\,G\,M(r)}{c^2\,r}\right)\,
\left(\frac{E^2}{c^2\,\alpha^2(r)}-\frac{L^2}{r^2}-1\right).
\label{eq:potes}
\end{equation}

Restricting the radial motion to circular stable orbits, implies
imposing the conditions, $\dot{r}=0$, for circular orbits, and
$\frac{\partial V}{\partial r}=0$, so that it describes an
extremum of the motion, and $\frac{\partial^2 V}{\partial r^2}>0$,
so that the extremum is a minimum. These three conditions
guarantee that the motion will be circular and stable. They also
imply the following expressions for the energy and total momentum
of the particles in such orbits:
\begin{eqnarray}
E^2 &=&{\frac{{c^2\,\alpha^3(r)}}{{\alpha(r)-r\,\alpha(r)_r}}}, \\
L^2 &=&{\frac{{r^3\,\alpha(r)_r}}{{\alpha(r)-r\,\alpha(r)_r}}},
\label{eq:ELes}
\end{eqnarray}
\noindent where a subindex $r$\ stands for derivative with respect
to $r$.

On the other hand, we can rewrite the line element for this
geometry, Eq.(\ref{eq:els}), in terms of the modulus of the
spatial velocity, normalized with the speed of light, measured by
an inertial observer far from the source, as
$ds^2=-dt^2\left(1-v^2\right)$, where
\be v^2=\frac{1}{c^2\,\alpha^2(r)}\, \left(
\frac{\left(\frac{dr}{dt}\right)^2}{1-\frac{2\,G\,M(r)}{c^2\,r}} +
r^2\,\left(\frac{d\Omega}{dt}\right)^2 \right). \ee
This last equation implies that the modulus of the angular
velocity, which is the tangential velocity for the case of
circular orbits, is defined as:
\be
{v_{\mathrm{tg}}}^2=\frac{r^2}{c^2\,\alpha^2(r)}\,\left(\frac{d\Omega}{dt}\right)^2
=\frac{1}{c^2\,\alpha^2(r)}\,\left(\frac{d\tau}{dt}\right)^2\,{\dot\Omega}^2,
\ee
thus, in terms of the conserved quantities, the angular velocity
takes the form:
\be
{v_{\mathrm{tg}}}^2=\frac{c^2\,\alpha^2(r)}{r^2}\,\frac{L^2}{E^2}.
\ee

Using the expression derived for these conserved quantities,
Eq.(\ref{eq:ELes}), we obtain that the tangential velocity can be
expressed in terms of the metric coefficient $\alpha$ as:
\begin{equation}
{v_{\mathrm{tg}}}^2=\frac{r\,\alpha(r)_r}{\alpha(r)}.
\label{eq:velt}
\end{equation}

This last equation allows us to determine the metric coefficient
$\alpha(r)$ in terms of the observed velocity profile:
\begin{equation}
\alpha(r)=\mathrm{exp} \int\frac{v_{\mathrm{tg}}^2(r)}{r}\,dr.
\label{eq:alph}
\end{equation}

This is the key equation of the reasoning: To use the
observations, $v_{\mathrm{tg}}(r)$, in order to partially
determine the geometry of the surrounding spacetime.

Now we join these results with the Navarro-Frenk-White model. This
model predicts the density profile \cite{NFW_mod}:
\begin{equation}
\rho _{\rm NFW}=\frac{\rho _0}{\frac{r}{r_s}(1 +
\frac{r}{r_s})^2}, \label{eq:d_nfw}
\end{equation}
where $\rho _0=\rho _{\rm crit}\,\delta _{c}$, $r_s$ is a scale
radius, $\delta_c$ is a characteristic (dimensionless) density,
and $\rho _{\rm crit}=\frac{3H^2}{8\,\pi\,G}$ is the critical
density for closure. The mass function, $M_{\rm
NFW}(r)=4\,\pi\,\int\,r^2\rho_{\rm NFW}\,dr$, with the integration
constant chosen so that $M_{\rm NFW}(r=0)=0$, takes the form:
\be M_{\rm NFW}(r)=4\,\pi\,r_s^3\,\rho_0\,\left(
\ln\left(1+\frac{r}{r_s}\right)-
\frac{\frac{r}{r_s}}{1+\frac{r}{r_s}}
 \right). \label{eq:M_nfw}\ee
This implies, equating the gravitational force with the
centrifugal one, the following profile for the tangential
velocity:
\begin{equation}
{v_{\mathrm{tg}}^2}_{NFW}=v_0^2\left(\frac{\ln
(1+\frac{r}{r_s})}{\frac{r}{r_s}} -
\frac{1}{1+\frac{r}{r_s}}\right) , \label{v_nfw}
\end{equation}
where $v_0^2=4\,\pi\,G\,\rho _0\,r_s^2$ is three times the
tangential velocity of particles at the limb of a sphere of radius
$r_s$ and a mass defined by the critical density times the
$\delta_c$ factor inside that volume.

These expressions are directly predicted by the NFW model and,
with exception of the central parts of galaxies, it has been
successfully compared with the actual observations \cite{NFW_mod}.

\begin{figure}[htb]
\includegraphics[width=5cm]{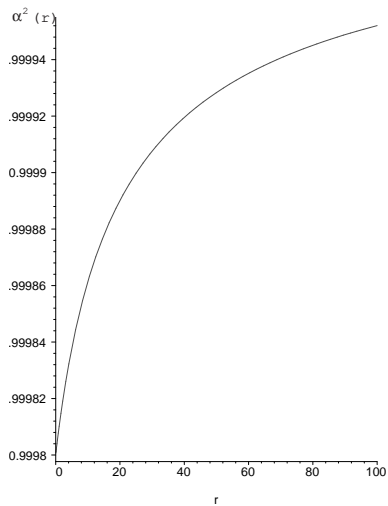}
\caption{\label{fig:alpha} The $g_{tt}$ metric coefficient, for
values $\frac{v_0^2}{c^2}=10^{-4} $, $r_s=10 kpc$.}
\end{figure}

Using the expression derived for the tangential velocity within
the NFW model, Eq. (\ref{v_nfw}), in the expression we obtained
above, Eq. (\ref{eq:velt}), we obtain a remarkable simple
expression for the $g_{tt}$ coefficient:
\begin{equation}
\alpha^2(r)=\left(1+\frac{r}{r_s}\right)^{-2\frac{v_0^2\,r_s}{c^2\,r}},
\label{eq:alpha}
\end{equation}
where we have set the integration constant such that $\alpha$ goes
to one for large radii, and we have normalized with the speed of
light the NFW velocity. There are several noticeable features in
this last expression. First, it is regular everywhere. The
divergency problem that the NFW-density has in the central region,
is not reflected in the metric coefficient:
\be \lim_{r\to 0}\alpha^2= e^{-\frac{v_0^2}{c^2}}, \\\\
\lim_{r\to \infty}\alpha^2=1. \ee
Actually, the $\alpha$-function goes to one for large radii, as
can be seen in the figure, \ref{fig:alpha}, recovering and
validating the Newtonian assumption in that region.

We can go on and work with the other metric coefficient,
$g_{rr}=(1-\frac{2\,G\,M(r)}{c^2\,r})^{-1}$. First of all, recall
that the $g_{tt}$ coefficient was determined based on the analysis
presented above. For the $g_{rr}$ coefficient, the unknown
function $M(r)$ can not be directly identified with the mass
function. Strictly speaking, the mass is defined for
asymptotically flat spacetimes with several characteristics, but
we do not go into that discussion in this work. We can say that
this is an approximation. On the other hand, however, as will be
shown below, the ${}^t_t$ component of the Einstein equations,
implies a consistency between the fact of taking this function as
the mass and the defined density, so that we have some grounds to
consider it like the mass. Thus, let us be bold and take the
function $M(r)$ as the mass function of the NFW model. The line
element takes the form:
\be
ds^2=-\left(1+\frac{r}{r_s}\right)^{-2\frac{v_0^2\,r_s}{c^2\,r}}\,c^2\,dt^2
+ \frac{dr^2}{1- 2\,\frac{v_0^2}{c^2}\,\left(
\frac{\ln\left(1+\frac{r}{r_s}\right)}{\frac{r}{r_s}}-
\frac{1}{1+\frac{r}{r_s}}
 \right)}+ r^2\,d\Omega^2,
\label{eq:elnfw} \ee

In this way, we obtain the geometry associated with the
Navarro-Frenk-White model.

The line element has two free parameters, namely $v_0$, the
characteristic speed, and $r_s$, the characteristic radius.

\begin{figure}[htb]
\includegraphics[width=5cm]{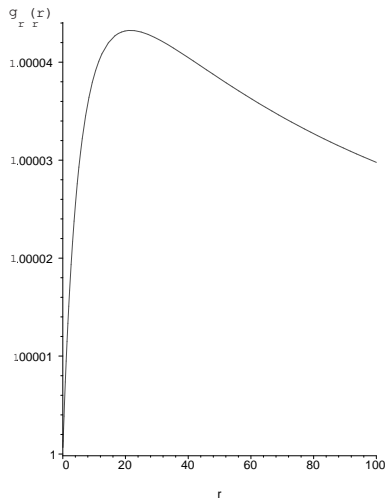}
\caption{\label{fig:grr} The $g_{tt}$ metric coefficient, for
values $\frac{v_0^2}{c^2}=10^{-4} $, $r_s=10 kpc$.}
\end{figure}

With respect to the $g_{rr}$ metric coefficient, it is also
regular everywhere, for positive radius, we see its behavior in
figure Fig.(\ref{fig:grr}), and it has the following limits:
\be \lim_{r\to 0}g_{rr}= \lim_{r\to \infty}g_{rr}=1. \ee

Thus, the line element given by Eq.(\ref{eq:elnfw}) is regular at
all points and far from the source takes the form of the flat
spacetime.

However, even though the line element does not show any
divergence, we expect to have one, directly inherited from the
divergence of the NFW density at the center. Actually, when one
derives the scalar of curvature, $R$, it is seen that:
\be R_{r\to 0} \sim
2\frac{v_0^2}{c^2\,r_s\,r}-2\frac{v_0^2}{c^2\,r_s^2}\left(2-\frac{v_0^2}{c^2}\right)
+ {\cal O}(r), \ee
showing a clear divergence at the center, as expected. This
implies that we do not have a solution describing a spacetime, as
long as there is a naked singularity, the divergence is not
covered by any horizon. What we have is a geometry which implies a
dynamic similar to the one described by the NFW model, so we claim
that this geometry will be related with the exterior part of a
complete spacetime with the observed dynamics.

The metric coefficients differ from the unity by very tiny
amounts, validating the flat space approximation, that is, the
Newtonian analysis. However, within the General relativistic
theory, these tiny amounts of non-flat geometry, are the ones
responsible for the observed dynamics!

Finally, we can construct the Einstein tensor and obtain some
conclusions about the type of matter-energy which produce the
space time given by Eq.(\ref{eq:elnfw}). The Einstein tensor gives
three non-zero independent components:
\bea
G^t_t&=&-2\frac{v_0^2}{c^2\,r_s^2}\,\frac{1}{x\,u^2},\label{eq:Gtt}
\\
G^r_r&=&-4\frac{v_0^4}{c^4\,r_s^2}\,\frac{1}{x^4\,u^2}\,
\left[u\,\ln u \,\left( u\,\ln u -2\,x \right) + x^2 \right],
\\
G^\theta_\theta&=&-\frac{v_0^4}{c^4\,r_s^2}\,\frac{1}{x^5\,u^3}\,
\left(u\,\ln u - x \right)\,\left[2\,u\,\ln
u\,\left(\frac{v_0^2}{c^2}\,u\,\ln u - 2\,x\,\left(u +
\frac{v_0^2}{c^2} \right) \right) +
x^2\,\left(7\,x+4+2\,\frac{v_0^2}{c^2}\right)\right], \eea
where we have defined $x=\frac{r}{r_s}$, and $u=1+\frac{r}{r_s}$.

Being aware of the caveats on promoting this geometry to a
spacetime, still we can say something about the type of matter
which could produce such a geometry, by means of the Einstein
equations:
\be G^\mu_\nu=\frac{8\,\pi\,G}{c^4}\,T^\mu_\nu, \ee
where $G$ stands for the gravitational constant and $T^\mu_\nu$
describes the tensor of distribution of the matter-energy in the
spacetime.

It is common to identify the $T^t_t$ component of the
matter-energy tensor with the density of matter, and energy,
present in the spacetime, $\rho$, that is $T^t_t=-c^2\,\rho$.
Thus, using the corresponding expression for the Einstein tensor,
Eq.(\ref{eq:Gtt}), we obtain the same expression relating the mass
and the density as that obtained in the Newtonian theory,
Eq.(\ref{eq:M_nfw}). This fact gives support to the interpretation
of the function $M(r)$ in the line element, Eq.(\ref{eq:elnfw}),
as the mass of the system. These are good news and give
consistency to the treatment presented here.

About the other components of the matter-energy tensor, by means
of the Einstein equations, we can conclude that such matter-energy
tensor can not be dust or even perfect fluid. The reason for these
conclusions are clear. Again, if we take the matter-energy tensor
to be a perfect fluid one
\be
T^\mu_\nu=\left(\rho\,c^2+p\right)u^\mu\,u_\nu+p\,\delta^\mu_\nu,
\ee
with $u^\mu$ the four velocity. For the spherically symmetric case
we are dealing with, the matter-energy tensor takes the form
$T^\mu_\nu={\rm diag}\left(-\rho\,c^2,p,p,p\right)$. But, from the
Einstein tensor, we see that clearly $G^r_r$, and
$G^\theta_\theta$ are non zero and different. Thus, the
matter-energy curving the spacetime to form the NFW geometry can
not be dust or perfect fluid.

If we insist on a dark fluid type for the matter-energy, it has to
be an anisotropic one, with two different pressures, a radial one,
$p_r$, and a tangential one, $p_\bot$ which, by the Einstein
equations imply
\bea p_r&=& -\rho^2\,2\,\pi\,G\,\frac{r_s^2\,u^2}{x^2}\,
\left[u\,\ln u \,\left( u\,\ln u -2\,x \right) + x^2 \right]\\
p_\bot&=& -\rho^2\,8\,\pi\,G\,\frac{r_s^2\,u}{x^3}\, \left(u\,\ln
u - x \right)\,\left[2\,u\,\ln u\,\left(\frac{v_0^2}{c^2}\,u\,\ln
u - 2\,x\,\left(u + \frac{v_0^2}{c^2} \right) \right) +
x^2\,\left(7\,x+4+2\,\frac{v_0^2}{c^2}\right)\right]. \eea

It is important to notice that, as the density, $\rho$ is very
small, these pressures are even smaller, and tend very quickly to
zero. Thus, again validate the Newtonian treatment taken the fluid
as dust, in an exterior region. Nevertheless, recall that these
analysis is on the track of determining the actual nature of the
dark matter, pointing to the physical properties it must have.

In this way, the geometry associated with the NFW model that we
have derived, allows us to explain the validity of the Newtonian
treatment, pointing to its limits, to discuss on the nature of the
dark matter, and opens the road to perform several dynamical
studies within the general relativistic theory.


We are grateful with Elloy Ayon for fruitful discussions during
the elaboration of the present work. DN acknowledges the
DGAPA-UNAM grant IN-122002.



\begin{thebibliography}{99}
\bibitem{larg}T. Matos, D. N\'u\~nez, F. S. Guzm\'{a}n and E. Ram\'{\i}rez,
Gen. Rel. and Grav. {\bf 34}, 283 (2002).
\bibitem{quint} F. S. Guzm\'{a}n, T. Matos, D. N\'u\~nez, and E. Ram\'{\i}rez,
Rev. Mex. de Fis. to appear, (2003).
\bibitem{NFW_mod}J. F. Navarro, C. S. Frenk, and S. D. M. White,
Astrop. J. {\bf 490}, 493, (1997).
\end{thebibliography}
\end{document}